\documentclass[twocolumn,superscriptaddress,10pt,floatfix]{revtex4-1}
\usepackage[utf8]{inputenc}
\usepackage{amsmath}
\usepackage{amssymb}
\usepackage{graphicx}
\usepackage{subfigure}
\usepackage{dcolumn}
\usepackage{enumerate}
\usepackage{textcomp}
\newcommand{\QE}{\mbox{Quantum Espresso}}
\newcommand{\Elk}{\mbox{Elk}}
\newcommand{\cminv}{$\textrm{cm}^{-1}$}
\newcommand{\mkm}{\textmu{}m}
\newcommand{\mks}{\textmu{}s}
\newcommand{\Biz}{$\textrm{Bi}^{0}$}
\newcommand{\Bipi}{$\textrm{Bi}^{+}$}
\newcommand{\Bipiii}{$\textrm{Bi}^{3+}$}
\newcommand{\AgCl}{\mbox{$\textrm{AgCl}$}}
\newcommand{\AgClBi}{\mbox{$\textrm{AgCl:Bi}$}}
\newcommand{\Term}[4]{\mbox{${}^{#1}{\textrm{#2}}_{#3}{#4}$}}
\newcommand{\Cub}[1]{\mbox{{#1}m$\overline{3}$m}}
\begin{document}
\title{%
Near-IR luminescence in bismuth-doped AgCl crystals
}
\author{V.~G.~Plotnichenko}
\affiliation{Fiber Optics Research Center of the Russian Academy of
Sciences \\ 38 Vavilov Street, Moscow, 119333, Russia
}
\affiliation{Moscow Institute of Physics and Technology \\
9 Institutskii per., Dolgoprudny, Moscow Region, 141700, Russia
}
\author{D.~V.~Philippovskiy}
\affiliation{Fiber Optics Research Center of the Russian Academy of
Sciences \\ 38 Vavilov Street, Moscow, 119333, Russia
}
\author{V.~O.~Sokolov}\email{Corresponding author: vence.s@gmail.com}
\affiliation{Fiber Optics Research Center of the Russian Academy of
Sciences \\ 38 Vavilov Street, Moscow, 119333, Russia
}
\author{V.~F.~Golovanov}
\affiliation{State Scientific-Research and Design Institute of Rare-Metal
Industry "Giredmet" JSC \\
5-1 B.Tolmachevsky lane, Moscow, 119017, Russia
}
\author{G.~V.~Polyakova}
\affiliation{State Scientific-Research and Design Institute of Rare-Metal
Industry "Giredmet" JSC \\
5-1 B.Tolmachevsky lane, Moscow, 119017, Russia
}
\author{I.~S.~Lisitsky}
\affiliation{State Scientific-Research and Design Institute of Rare-Metal
Industry "Giredmet" JSC \\
5-1 B.Tolmachevsky lane, Moscow, 119017, Russia
}
\author{E.~M.~Dianov}
\affiliation{Fiber Optics Research Center of the Russian Academy of
Sciences \\ 38 Vavilov Street, Moscow, 119333, Russia
}

\begin{abstract}
Experimental and computer-modeling studies of spectral properties of crystalline
\AgCl{} doped with metal bismuth or bismuth chloride are performed. Broad
near-IR luminescence band in the 0.8--1.2~\mkm{} range with time dependence
described by two exponential components corresponding to the lifetimes of 1.5
and 10.3~\mks{} is excited mainly by 0.39--0.44~\mkm{} radiation. Computer
modeling of probable Bi-related centers in \AgCl{} lattice is performed. On the
basis of experimental and calculation data a conclusion is drawn that the IR
luminescence can be caused by \Bipi{} ion centers substituted for Ag$^+$ ions.
\end{abstract}
\maketitle

During the last decade bismuth-doped bulk glasses and optical fibers are
extensively investigated due to their characteristic broadband IR luminescence
in the 1.0--1.7~\mkm{} range, enabling one to develop optical lasers and
amplifiers for fiber-optic communication systems (e.g., see review
\cite{Dianov13}). However the origin of bismuth-related IR luminescence centers
is not established yet, thus impeding the development of this field. Recently
(see, for example, discussion in Ref.~\cite{Peng11}) an opinion is strengthened
that the IR luminescence is caused by bismuth in subvalent states, and first of
all in monovalent state (\Bipi). To study the optical characteristics of
bismuth-related centers of such a nature it is convenient to use crystalline
halides of monovalent elements as a modeling host with a simple structure
(primitive, \Cub{P}, or face-centered, \Cub{F}, cubic lattices), in which the
impurity bismuth should form substitutional centers just in monovalent state. 

The IR luminescence in bismuth-doped crystalline halides of monovalent metals
was found and investigated for the first time in \mbox{CsI:Bi} (primitive
\Cub{P}{} lattice), and possible structure of the centers responsible for the
luminescence were suggested \cite{Su11, Su12}. Later the IR luminescence was
observed in \mbox{TlCl:Bi} crystals (\Cub{P}{} lattice) \cite{We13a}. Probable
IR luminescence centers in \mbox{CsI} and \mbox{TlCl} were studied then by
computer simulation technique \cite{We13a, We13b}.

The present paper is devoted to the study of IR luminescence in \AgClBi{}
crystals with face-centered \Cub{F}{} lattice.

\begin{figure}
\includegraphics[width=8.50cm, bb=70 280 545 770]{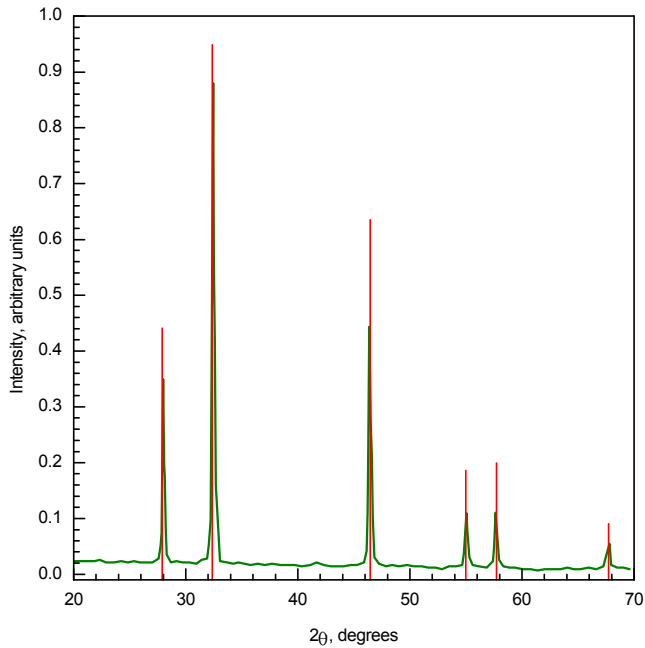}
\caption{%
XRD pattern of \AgCl{} doped with bismuth chloride. Vertical lines correspond to
the the standard XRD pattern of \AgCl{} (JCPDS No.~85-1355,
lattice constant 0.5549~nm).
}
\label{fig:XRD}
\end{figure}

The crystal was grown from silver chloride salts deposited from aqueous AgNO$_3$
solution after drying and purification by the method of directional (oriented)
melt crystallization. Before growth 30~g of \AgCl{} ligature with metal bismuth
(Bi) or bismuth chloride (BiCl$_3$) dopant (0.076~mass.\%{} of Bi) were prepared. 
Ligature was sealed in vacuum inside 16~mm-diameter ampule and melted at 
400--450~$^\circ$C. The melt was mixed no less than 5--6 times by the method of 
ampule reverse (flip-chip) and cured for 3 days. After cooling to room temperature 
the ligature was overloaded into the growth ampule 18~mm in diameter, supplied 
with pure silver chloride to total 100~g, melted in vacuum, mixed 3--5 times again 
and grown up by Bridgman-Stockbarger method at a rate of 1~mm/h. Growth was
proceeded for about 7 days. After growth the crystal was quenched to room
temperature at a rate of 50--60~$^\circ$C/h. BiCl$_3$-doped samples have
appeared to be dim or opaque. In samples doped with metal bismuth the inclusions
in the form of metal flakes (presumably silver) were observed. For spectroscopic
measurements two samples of 18~mm diameter and 5~mm thickness were cut from the
grown crystals, one from initial part of the crystal (where crystallization
starts during crystal growth), and the other from the final one (where
crystallization ends). X-ray diffraction (XRD) analysis of powders of all the
samples was performed on DRON~3M diffractometer using filtered Cu
$\mathrm{K}_\alpha$ radiation. JCPDS database was used to identify crystalline
phases. The XRD measurements revealed diffraction patterns typical for
rock-salt-type \AgCl{} crystal (Fig.~\ref{fig:XRD}). Lattice constant was found
to be $0.5596(1)$~nm in metal-doped \AgCl{} and $0.5574(1)$~nm in BiCl$_3$-doped
\AgCl{} (to be compared with $0.5597(1)$~nm in reference non-doped \AgCl{}
crystal).

\begin{figure}
\includegraphics[width=8.50cm, bb=70 280 545 770]{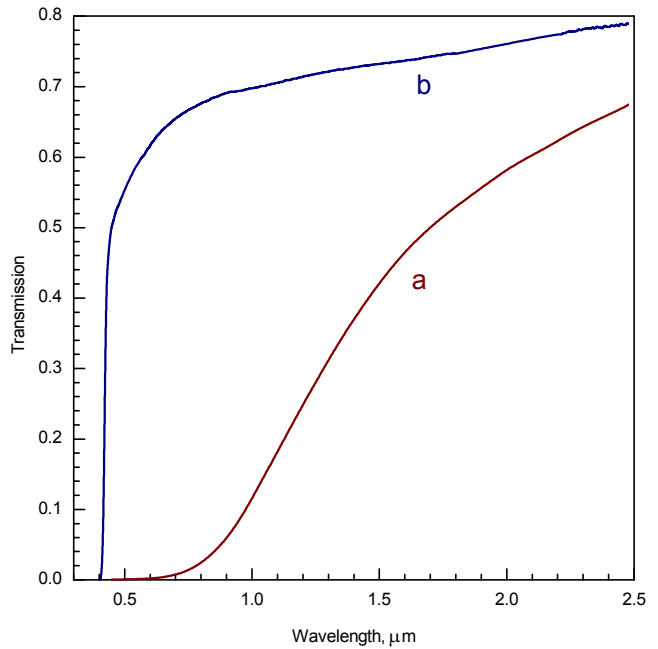}
\caption{%
Transmission spectra: (a)~\AgCl{} doped with bismuth chloride; (b)~\AgCl{} doped
with metal bismuth.
}
\label{fig:Transmission}
\end{figure}

The transmission spectra were measured on Perkin-Elmer Lambda~900 spectrometer,
and the luminescence spectra on Edinburgh Instruments FLSP920 spectrometer. The
transmission spectra of two samples are shown in Fig.~\ref{fig:Transmission}.
The short-wave transmission edge of \AgCl{} sample doped with metal bismuth is
near 0.41~\mkm{} that fits well with the known experimental data for pure
\AgCl{} (transmission edge is near 0.42~\mkm{} \cite{Weber03}, indirect energy
gap is about 3.2~eV \cite{Tejeda75}). In \AgCl{} samples doped with bismuth
chloride a considerable decrease in transmission is observed in the near IR
range. In the spectra of both samples there are no distinctly expressed
absorption bands.

\begin{figure}
\includegraphics[width=8.50cm, bb=70 280 545 810]{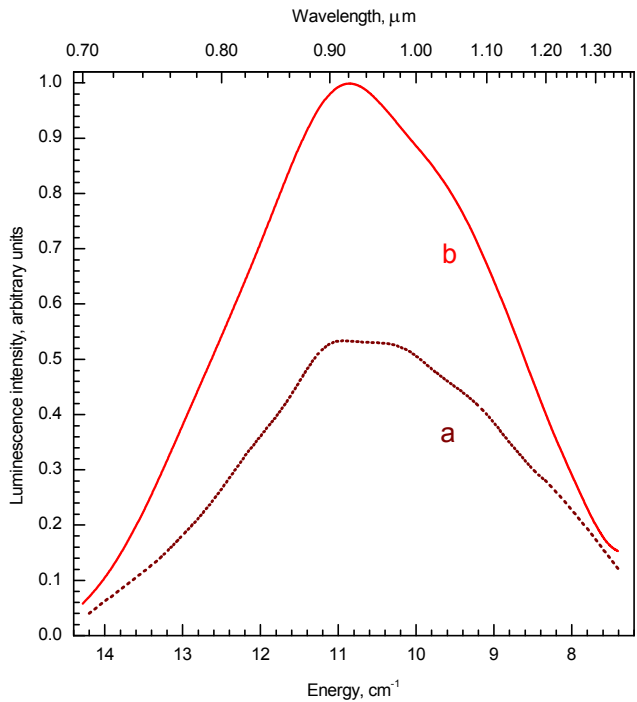}
\caption{%
Luminescence spectra excited at 0.42~\mkm{} wavelength: (a)~\AgCl{} doped with
bismuth chloride; (b)~\AgCl{} doped with metal bismuth.
}
\label{fig:Luminescence}
\end{figure}

\begin{figure}
\includegraphics[width=8.50cm, bb=70 280 545 810]{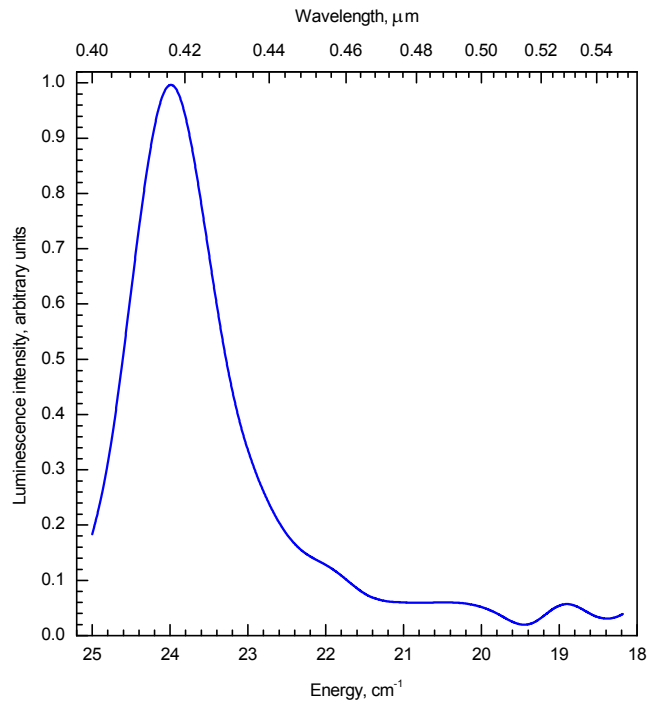}
\caption{%
Excitation spectrum of 0.90~\mkm{} luminescence in \AgCl{} sample doped
with metal bismuth.
}
\label{fig:Excitation}
\end{figure}

In the luminescence spectra a wide IR luminescence band is discovered with a
maximum near 0.90~\mkm{} and width about 0.30~\mkm{}
(Fig.~\ref{fig:Luminescence}). The luminescence in this band is excited mainly
in the absorption band with a maximum near 0.42~\mkm{}
(Fig.~\ref{fig:Excitation}). The efficiency of luminescence excitation at other
wavelengths is much lower. No luminescence is found in AgCl samples containing
no bismuth.

The intensity of IR luminescence in \AgCl{} samples doped with metal bismuth is
considerably (about two times) higher in comparison with the samples doped with
bismuth chloride. On the other hand, in metal-doped \AgCl{} samples the
luminescence intensity noticeably changes along the grown crystal length: the
sample cut from the final part exhibits much more intensive luminescence. This
fact could be explained by bismuth displacement together with the moving
crystallization zone resulting in the content of active bismuth centers
increasing from initial part to the final part of the produced crystal.

The suppression of IR luminescence from measurement to measurement is found out:
with each new irradiation of the sample near the main excitation band, i.e.
within 0.39--0.44~\mkm{} range, the luminescence intensity decreases. At the
same time, the shape of the luminescence spectrum actually does not change. At
excitation in a longer-wave range, e.g. near 0.53~\mkm{}
(Fig.~\ref{fig:Excitation}), the effect of degradation is practically absent;
however the efficiency of luminescence excitation is extremely low.

Time dependence of IR luminescence is described by two exponential components
$a_1\exp\left(-t/\tau_1\right) + a_2\exp\left(-t/\tau_2\right)$, where $a_1
\approx 0.2, a_2 \approx 0.8$ (in arbitrary units), $\tau_1 \approx 10.4$~\mks,
$\tau_2 \approx 1.5$~\mks. Within the fitting accuracy (about 10\%) these
parameters do not differ in metal-doped and chloride-doped samples. The
luminescence lifetimes are considerably shorter than those in \mbox{CsI:Bi} (130
and 210~\mks{} \cite{Su11, Su12}) and in \mbox{TlCl:Bi} (200--350~\mks{}
\cite{We13a}) crystals, presumably due to high degree of disorder typical for
cation sublattice in \AgCl{} crystals.

Transmission decrease in the near IR range in BiCl$_3$-doped \AgCl{} samples can
be explained by the fact that for such a doping technique bismuth is
incorporated in the crystal lattice mainly in trivalent state (\Bipiii),
substituting for silver ion. This substitution should be accompanied by a
formation of two cation vacancies compensating an excess charge of
substitutional center. Thus, instead of three sites with monovalent cations
there are: a site with trivalent cation, two cation vacancies and two monovalent
interstitial Ag$^+$ ions (in other words, two Frenkel pairs). Similar lattice
disordering can be followed by an increase both in scattering  and absorption
near the short-wave edge as a result of occurrence of localized states in the
forbidden zone. In the case of metal doping, bismuth is most likely introduced
in the crystal lattice mainly as monovalent substitutional impurity (\Bipi),
without causing an appreciable lattice disordering, and accordingly, a
significant absorption or scattering. Rather low intensity of IR luminescence of
BiCl$_3$-doped \AgCl{} samples in comparison with metal-doped \AgCl{} samples
can be presumably explained just by considerable optical losses in the
first-type samples.

To understand the origin of bismuth-related centers responsible for the revealed
IR luminescence, we performed computer simulation of the structure and
absorption spectra of several most probable centers which could be formed by
impurity bismuth atoms in \AgCl{} crystal lattice, namely: monovalent
substitutional \Bipi{} centers, complexes formed by substitutional bismuth
center in cation site and silver or chlorine vacancy in the nearest cation or
anion lattice site, and interstitial \Bipi{} ions. For the modeling we used
$2\times2\times2$ or $2\times2\times3$ \AgCl{} supercells containing 64 or 96
atoms, accordingly. In the central part of supercell silver atom was substituted
by bismuth atom, an interstitial bismuth atom was introduced, and also vacancies
were formed by a removal of one of silver or chlorine atoms. The charge state of
the center was given by a total supercell charge. Equilibrium configurations of
bismuth-related centers were found by a complete optimization of geometry with
the gradient method in the plane wave basis within the generalized gradient
approximation of density functional theory using pseudopotentials. Such
calculations were carried out by means of \QE{} software package \cite{QE}. Thus
obtained configurations of bismuth-related centers were used further to
calculate the absorption spectra of the centers by Bethe-Salpeter equation
method based on all-electron full-potential linearized augmented plane wave
approach taking into account spin-orbit interaction essential for
bismuth-containing systems. For calculations the \Elk{} code \cite{Elk} was
used. To estimate the Stokes luminescence shift, the configurational coordinate
diagrams of bismuth-related centers in the frame of a simple model were
calculated as well. The obtained results show that for all bismuth-related
centers studied there are bases to consider the Stokes shift small and to
estimate the luminescence wavelength without taking the shift into account.
Detailed description of the calculation technique is given in Refs.~\cite{We13a,
We13b}.

Calculation results on bismuth-related centers are convenient for interpreting
within the limits of the models used earlier for bismuth-related centers in
\mbox{TlCl:Bi}, \mbox{CsI:Bi}, \cite{We13a}, \mbox{SiO$_2$}, and \mbox{GeO$_2$}
\cite{We13b}. As is known, the ground and the first two excited states of
\Bipi{} ion arise from \Term{3}{P}{}{}{} triplet (an electronic configuration
6s$^2$p$^2$) split by strong spin-orbit interaction into components:
\Term{3}{P}{0}{}{} (ground state), \Term{3}{P}{1}{}{} and \Term{3}{P}{2}{}{} 
(excited states with the energies for a free ion near 13300 and 17000~\cminv,
accordingly). These states can be split and mixed under the influence of the
environment of the \Bipi{} ion in the crystal lattice.

\begin{figure}
\includegraphics[height=13.75cm, bb=130 95 480 697]{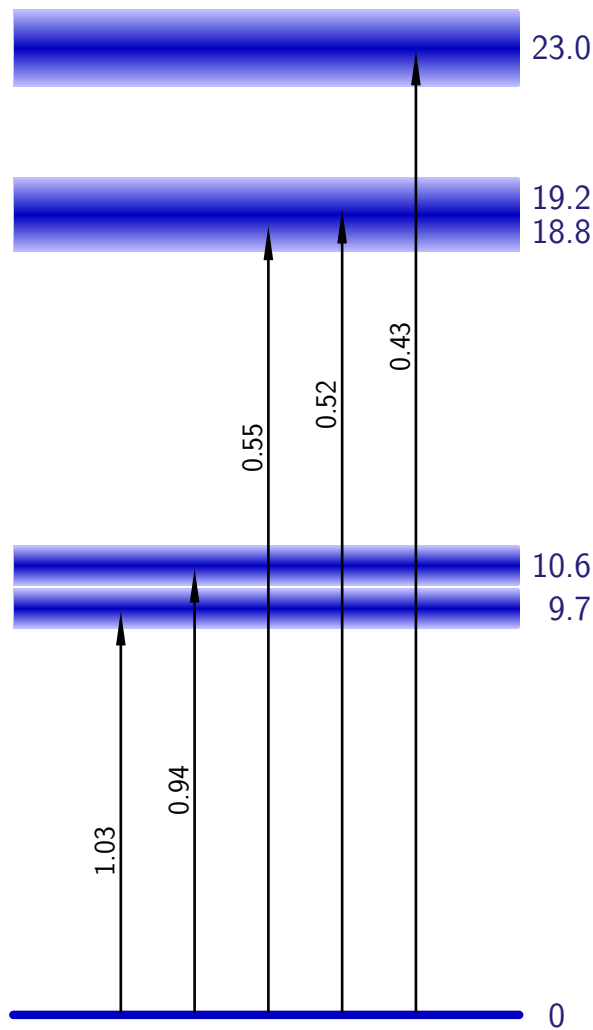}
\caption{%
Calculated levels and transitions of impurity substitutional \Bipi{} center
(energy levels in $10^3$~\cminv, transition wavelengths in \mkm).
}
\label{fig:Levels}
\end{figure}

In the case of monovalent substitutional \Bipi{} center such a transformation of
states is due to a redistribution of electronic density between the bismuth ion
and the neighboring chlorine ions, but is not due to the crystal field (owing to
a cubic symmetry of the substitutional center). According to the calculations,
in \AgClBi{} such a redistribution appears to be rather weak. Energy levels and
transitions between them calculated for monovalent substitutional bismuth center
\Bipi{} are shown in Fig.~\ref{fig:Levels}. It is worth noticing that splitting
the first excited state \Term{3}{P}{1}{}{} is insignificant. Considering in
accordance with the above-said the Stokes shift to be small, one could expect
for monovalent substitutional bismuth center, \Bipi, the occurrence of IR
luminescence in the 0.9--1.0~\mkm{} region excited in two absorption bands near
0.4 and 0.5~\mkm.

Calculation results have shown that the influence of \AgCl{} lattice on other
bismuth-related centers under study is substantially reduced to the crystal
field effect and in a qualitative sense is described by the model similar to
that suggested in Ref.~\cite{Mollenauer83} and used in Refs.~\cite{We13a,
We13b}. However, we should emphasize that although this model allows visual
interpreting of the origin of states in bismuth-related centers, it describes
the occurring electronic structure only in the rough. The quantitative
conclusions are based on the performed calculation results.

In the case of interstitial \Bipi{} ion the crystal field appears to be too weak
and causes only insignificant splitting and shift of ion states. Therefore the
interstitial \Bipi{} ion in \AgCl, like \Bipi{} ions in chloride solutions
\cite{Davis67}, can cause a weak absorption in the bands near 0.7, 0.6 and
0.5~\mkm, accompanied by a luminescence near 0.7~\mkm. The experimental data
available do not allow to make certain conclusions about the presence or absence
of interstitial \Bipi{} ions in \AgClBi.

On the contrary, in the complex formed by substitutional bismuth center and
chlorine vacancy in the nearest lattice site, the crystal field appears to be
strong. Thus there is a redistribution of electronic density between bismuth
atom and chlorine vacancy, so that in rough approximation such a complex
represents a pair of charged centers: \Bipi{} ion in cation site and negatively
charged chlorine vacancy. In axial crystal field influenced by the vacancy, the
ground state of Bi+ ion (\Term{3}{P}{0}{}) is not split at all, the excited
\Term{3}{P}{1}{}{} state is split into two levels with the energies about
4400 and 10500~\cminv, and the excited \Term{3}{P}{2}{}{}  state is split into
three levels with the energies about 12000, 17000 and 22000~\cminv. Considering
the Stokes shift to be small, it is possible to expect for the complex
“substitutional bismuth center --- chlorine vacancy” the IR luminescence in the
$>2.0$~\mkm{} region, and probably in the 1~\mkm{} region, excited in absorption
bands near 0.8, 0.6 and 0.4~\mkm. As follows from a comparison of these results
with the experimental data, similar complexes can make only an insignificant
contribution to the IR luminescence observed in \AgClBi.

As to the complex “substitutional bismuth center --- silver vacancy”, it was
found to be highly unstable in our calculations: owing to lattice
transformation in the vicinity of the complex, silver ion from the next
coordination shell fills the vacancy and so the complex actually disappears.

Thus, the performed modeling gives the grounds to believe that the main
contribution to the IR luminescence in \AgClBi{} is made by monovalent
substitutional bismuth centers, \Bipi. In this case an observable decrease in
intensity of the luminescence excited in absorption band near 0.42~\mkm{} can
be explained. Radiation with wavelength close to the edge of interband
absorption of \AgCl{} excites the electron-hole pairs. Capturing electrons from
the conductivity band, monovalent substitutional bismuth centers \Bipi{} are
restored to neutral \Biz{} centers, making no contribution to the IR
luminescence.

\section*{%
Acknowledgments
}
The authors are grateful to S.~V.~Firstov for assistance in luminescence
measurements.
This work is supported
in part by Basic Research Program of the Presidium of the Russian Academy of
Sciences and by Russian Foundation for Basic Research (grant
\mbox{12-02-00907}).
%
%

\end{document}